\documentclass[
aps,
amsmath,amssymb,
prl,
reprint,
groupedaddress,
superscriptaddress,
showpacs,
floatfix,
]{revtex4-1}

\usepackage{graphicx}
\usepackage{dcolumn}
\usepackage{bbm}
\usepackage{bm}
\usepackage{float}
\usepackage{natbib}
\usepackage{hyperref}
\usepackage{makecell} 
\usepackage{scalerel}
\hypersetup{colorlinks=true,citecolor=blue,urlcolor=blue,linkcolor=blue}

\begin{document}
\title{Passive and Deterministic Controlled-phase Gate for Single-photon Wavepackets Based on Time-reversal Symmetric Photon Transport}
\author{Zhaohua Tian}
\affiliation{School of Physics, Wuhan National Laboratory for Optoelectronics, Institute for quantum science and engineering and Hubei Key Laboratory of Gravitation and Quantum Physics, Huazhong University of Science and Technology, Luoyu Road 1037, Wuhan, 430074, China}

\author{Xue-Wen Chen}
\email[To whom all correspondence should be addressed:\\ ]{xuewen\_chen@hust.edu.cn}
\affiliation{School of Physics, Wuhan National Laboratory for Optoelectronics, Institute for quantum science and engineering and Hubei Key Laboratory of Gravitation and Quantum Physics, Huazhong University of Science and Technology, Luoyu Road 1037, Wuhan, 430074, China}
\affiliation{Wuhan Institute of Quantum Technology, Wuhan 430206, China}

\date{\today}

\begin{abstract}
We report the construction of a passive, deterministic and near-unity-fidelity controlled-$\pi$-phase gate for single-photon wavepackets with a node comprising a two-level emitter and a small number of cavities. The proposed gate is rooted in the concept of time-reversal symmetric photon transport, which makes the entire photon transport process into a perfect absorption and re-emission process. Consequently, it can circumvent the longstanding issue of wavepacket distortion inherent in photonic phase gates employing nonlinear media. Moreover, such time-reversal symmetric transport ensures the nonlinear $\pi$ phase shift by a single two-level emitter for the two-photon case. We develop analytical solutions to reveal the temporal dynamics of the nonlinear photon transport and to optimize the gate structure. Notably, the gate fidelity can exceed 99\% for a node with only four cavities for both single-photon and two-photon operations. Moreover, the proposed gate architecture is compatible with the platforms of integrated photonics. 
\end{abstract}
\maketitle

\par\emph{\textbf{Introduction.---}}
Single photons as flying qubits, in many aspects, are ideal for quantum information processing. However, the absence of direct interaction between photons poses a grand challenge in realizing logic gates like controlled-$\pi$-phase gate, which are needed for universal quantum computing. A number of schemes have been proposed to circumvent the challenge. A well-known approach is to use detection as effective nonlinearity \cite{KnillNature2001Linear, MilburnRevModPhys2007LinOpt}. But this scheme is probabilistic in nature and demands large resource overheads \cite{SimonPhysRevX2015}. Researchers have also proposed a variety of deterministic schemes to realize photon interactions indirectly via nonlinear media, such as single three-level systems in a cavity \cite{DuanPRL2004,YasyunobuPRA2010Swap,FanOptica2021,RitterNature2016,ZubairyPRA2003}, single four-level systems \cite{ZhengPRL2013} or a chain of three-level systems \cite{AndersPRL2022} in a waveguide, ensembles of atoms \cite{SorensenPRL2018}, Rydberg atoms \cite{LukinPRL2011Rydberg,RempeNatPhys2019Rydberg}, spin-chain systems \cite{LukinPRL2010Spin}, single two-level systems with quadratic coupling \cite{FelicettiPRXQuantum2023}, and so on. 
While these schemes offer tantalizing insights into physics and promising prospects, they typically only apply for photons at the single-frequency limit. In practice, single photons manifest as wavepackets with finite spectral linewidths, thus encountering wavepacket distortion following the gate operations. As pointed out by a number of researchers \cite{ShapiroPRA2006,ShapiroNJP2007,ShapiroPRA2014LimTLS,BanaclochePRA2010KerrLimit,FanPRL2013,DirkPRA2017LimitTLS}, for travelling photon wavepackets, it appears impossible to obtain high-fidelity phase gate based on Kerr nonlinearity and/or two-level emitters. 

To address the issue of wavepacket distortion, researchers have proposed dynamic modulation of the coupling between photons and nonlinear media \cite{DirkPRL2020,ZouPRApplied2020,HeucknpjQuan2022}. The dynamic modulation should be delicately designed and precisely implemented. Another class of approach is to use photon number sorting together with sum frequency generation and wavepacket reshaping \cite{LodahlPRL2015,KlausPRL2022}. Meanwhile, Brod et al. reported a passive scheme of using cross-Kerr interactions \cite{CombesPRL2016Passive}. For two-photon operation, the gate could achieve fidelity over 99\% with 12 cross-Kerr interaction sites and 24 identical atoms. Although the scheme is challenging for realization and the fidelity for single-photon operation is unsatisfactory, it is inspirational by showing the possibility to construct a high-fidelity phase gate via nonlinear media. 

Here we propose a near unity-fidelity, passive and deterministic controlled-$\pi$-phase (CZ) gate based on a node comprising one emitter and a small number of cavities. We intuitively elucidate the physics associated with the gate operation that is free of wavepacket distortion. With the development of an analytical solution to single-photon and two-photon transport dynamics, we provide the procedure to devise and optimize the gate. We also discuss the prospect of on-chip realization of the gate. 

\begin{figure}[!htbp]
	\centering
	\includegraphics[width=84mm]{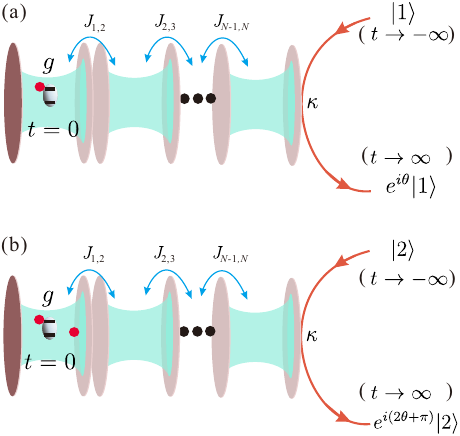}
	\caption{Schematics of the node architecture and the proposed CZ gate operation for (a) single-photon and (b) two-photon wavepackets, respectively. The red balls indicate the excitation quantas at the time-reversal symmetric point $t=0$.}
	\label{fig:fig1}
\end{figure}

\par \emph{\textbf{The physical picture.---}} A generic CZ gate operation can be expressed as \cite{NielsenBook2002quantum} $|a\rangle_{c}|b\rangle_{t} \longrightarrow \allowbreak e^{i \theta(a+b)+i \pi \mathrm{ab}}|a\rangle_{c}|b\rangle_{t}$, with $a,b=0,1$ denote the logic value of the control and target qubits. For dual-rail encoded photonic qubits, a CZ gate realizes
\begin{equation}\label{eq:inputctqubit}
	|1\rangle\rightarrow e^{i\theta}|1\rangle, \quad |2\rangle\rightarrow e^{i(2\theta+\pi)}|2\rangle,
\end{equation}
where $|1\rangle=|\phi_{1}\rangle_{\text{in}}=\int d\tau f_{\mathrm{I}}(\tau)\hat{c}_{\tau}^{\dagger}|\varnothing\rangle$ is a single-photon wavepacket and $|2\rangle=|\phi_{2}\rangle_{\mathrm{in}}=\iint\allowbreak d\tau_{1}d\tau_{2} \allowbreak f_{\mathrm{I}}(\tau_{1}) \allowbreak f_{\mathrm{I}}(\tau_{2})(\hat{c}_{\tau_{1}}^{\dagger}\hat{c}_{\tau_{2}}^{\dagger}/\sqrt{2})|\varnothing\rangle$ is a two-photon wavepacket, where $f_{\mathrm{I}}(\tau)$ satisfies the normalization condition $\int|f_{\mathrm{I}}(\tau)|^2d\tau=1$. Here $|\varnothing\rangle$ denotes the vacuum state and $\hat{c}_{\tau}^{\dagger}$ is the continuous-time creation operator \cite{BlowPRA1990ContOperator}. State $|2\rangle$ includes the control and target single photons, which are in the same mode and indistinguishable. $\theta$ is the phase that a single-photon wavepacket acquires after the gate operation. The challenges to establish a photonic CZ gate lie in the following two aspects, to induce a nonlinear $\pi$ phase for the two-photon case, and to keep the single- and two-photon wavepackets undistorted. 

\par Our CZ gate is illustrated in Fig. \ref{fig:fig1} with the central part being a node comprising a two-level emitter (TLE) and a few chain-coupled cavities, where the first cavity is coupled to the TLE and the last one interfaces with the input and output single-/two-photon wavepackets. The idea is to realize time-reversal symmetric photon wavepacket transport as depicted in Fig. \ref{fig:fig1}(b). By setting the system Hamiltonian $\hat{H}$ and the shape of the input wavepacket properly, one could realize a perfect absorption of the input wavepacket at time $t=0$. In particular, we require that at $t=0$, the single-photon input state  $|\phi_{1}(-t_{\infty})\rangle_{\text{in}}$ will only excite the TLE, while the two-photon input state $|\phi_{2}(-t_{\infty})\rangle_{\text{in}}$ will excite the TLE and its coupled cavity as indicated by the red balls in Fig. \ref{fig:fig1}. Here $-t_{\infty}$ denotes the time long before the interaction. For the two-photon case, since the TLE can only absorb one excitation, the other excitation absorbed by the cavity offers the asymmetry, which introduces an extra $\pi$ phase to the two-photon wavepacket at the output. We set the system state $|\Psi_{m}(t = 0)\rangle$ to be the symmetric point of the entire time-reversal symmetric transprot process. Here $m=1$ or $2$ denotes the single- or two-photon case. The output state can be considered as the result of emission from $|\Psi_{m}(0)\rangle$ while the input can be treated as the reversal of the absorption process \cite{TaylorBook2006Scattering,FanPRA2010InputOutput},
\begin{equation} \label{eq:phi_t1}
	|\phi_{m}\rangle_{\text{out}}=e^{-i\hat{H}t_{\infty}}\allowbreak|\Psi_{m}(0)\rangle,
	|\phi_{m}\rangle_{\text{in}}=e^{i\hat{H}t_{\infty}}\allowbreak|\Psi_{m}(0)\rangle.
\end{equation}
Here $\hbar = 1$ is used for convenience. The system Hamiltonian and $|\Psi_{m}(0)\rangle$ provide the flexibility to control the wavepackets and ensure high fidelity for both single- and two-photon cases. Thus, by realizing time-reversal symmetric photon transport, one could overcome the two aforementioned challenges of a photonic CZ gate.

\par \emph{\textbf{Analytic description of transport dynamics.---}} In the following, we present a rigorous theoretical description of photon transport dynamics with an emphasis on the two-photon case. The system Hamiltonian reads $\hat{H}=\hat{H}_{0}+\hat{H}_{I}$, 
where $\hat{H}_{0}=(\omega_{0}-i\Gamma_{0}/2) \hat{\sigma}_{+} \hat{\sigma}_{-}+\sum_{n=1}^{N} (\omega_{\text{C}}-i\Gamma_{\text{C}}/2) \hat{a}_{n}^{\dagger} \hat{a}_{n}+\int_{0}^{\infty} d \omega \omega \hat{c}_{\omega}^{\dagger} \hat{c}_{\omega}$ is the uncoupled Hamiltonian and the coupled one can be written as 
\begin{equation}	
	\begin{aligned}
		\hat{H}_{I}&=g\hat{\sigma}_{+}\hat{a}_{1}+\sum_{n=1}^{N-1} J_{n, n+1}\hat{a}_{n}^{\dagger} \hat{a}_{n+1}\\
		&+\sqrt{\frac{\kappa}{2 \pi}} \int_{0}^{\infty} d \omega \omega\hat{c}_{\omega}^{\dagger} \hat{a}_{N}
		+\text{h.c.}
	\end{aligned}
\end{equation}
Here $\omega_{0}$ is the transition of TLE with $\Gamma_{0}$ being its natural linewidth, and $\hat{\sigma}_{\pm}$ is its raising/lowering operator. The cavities are described by the same resonant frequency $\omega_{\text{C}}$, loss rate $\Gamma_{\text{C}}$ and the creation (annihilation) operators $\hat{a}_{n}^{\dagger}$ $(\hat{a}_{n})$. $g$ is the coupling constant between the TLE and the first cavity of the chain and $\kappa$ is the coupling rate between the last cavity of the chain and the waveguide. The neighboring cavities are coupled with a rate of $J_{n,n+1}$. The integral with the continuous-frequency creation (annihilation) operator $\hat{c}_{\omega}^{\dagger}$ $(\hat{c}_{\omega})$ represents the contribution from the waveguide continuum. $\hat{c}_{\omega}$ is related to the continuous-time annihilation operator $\hat{c}_{\tau}$ through the Fourier transformation. The abbreviation h.c. stands for Hermitian conjugate. Related system Hamiltonians have been studied in the context of waveguide QED \cite{KeijiPRA2004,FanPRL2007,FanPRA2009SinTransI,ShiPRB2009,LiaoJieQiaoPRA2010,ZhengPRL2013,ShiPRA2015MasterEquation,FirstenbergRMP2017,AndreaAluOptica2019,KlausPRL2019,AndesPRX2020,LiaoPRA2020,TianPRAppQST,YuxiLiuPRL2023,PoddubnyRMP2023}. We have applied the rotating-wave approximation. To grasp the main physics, in the following we consider the resonant case ($\omega_{0} = \omega_{\text{C}}$) and neglect the losses due to the spontaneous emission and cavity radiations, which are valid when $g\gg\Gamma_{0}$ and $J_{n,n+1},\kappa\gg\Gamma_{\text{C}}$. 

\par We adopt the wave-function approach \cite{StolyarovPRA2019,KlausPRL1992Wavefunctionapproach} to obtain the temporal dynamics of the system state. For single-photon transport, the system state can be described as $\left|\Psi_{1}(t)\right\rangle=\left[\sum_{n=0}^{N}c_{n}(t) \hat{a}_{n}^{\dagger}+\int_{0}^{\infty} d \omega c_{\omega}(t) \hat{c}_{\omega}^{\dagger}\right]|\oslash\rangle$. For two-photon transport, the state can be expressed as
\begin{equation}
	\begin{aligned}
		\left|\Psi_{2}(t)\right\rangle=& \left[\sum_{n=0}^{N} \sum_{m > n}^{N} c_{n,m}(t) \hat{a}_{n}^{\dagger} \hat{a}_{m}^{\dagger}+\sum_{n=1}^{N} c_{n,n}(t) \frac{\hat{a}_{n}^{\dagger} \hat{a}_{n}^{\dagger}}{\sqrt{2}} +\right.\\
		&+\sum_{n=0}^{N} \int_{0}^{\infty}d\omega c_{n,\omega}(t) \hat{a}_{n}^{\dagger} \hat{c}_{\omega}^{\dagger}\\
		&\left.+\iint_{0}^{\infty} d \omega_{1}  d \omega_{2} c_{\omega_{1},\omega_{2}}(t) \frac{\hat{c}_{\omega_{1}}^{\dagger} \hat{c}_{\omega_{2}}^{\dagger}}{\sqrt{2}}\right]|\oslash\rangle.
	\end{aligned} 
\end{equation}
We have used $\hat{a}_{0}/\hat{a}_{0}^{\dagger}$ to denote $\hat{\sigma}_{-}/\hat{\sigma}_{+}$ for convenience. The coefficient $c_{n,m}$ represents the complex probability amplitude of the state denoted by the subscripts. They can be classified into three groups, (i) the group with no excitation in the waveguide $\bm{c}_{\scaleto{\mathrm{XX}}{4pt}}=[c_{\scaleto{\mathrm{0,1}}{5pt}},c_{\scaleto{\mathrm{0,2}}{5pt}},...,c_{\scaleto{N,N}{5pt}}]^{\mathrm{T}}$, (ii) the group with one excitation in the waveguide $\bm{c}_{\scaleto{\mathrm{X\omega}}{5pt}}=[c_{\scaleto{\mathrm{0,\omega}}{5pt}},c_{\scaleto{\mathrm{1,\omega}}{5pt}}\allowbreak,...,\allowbreak c_{\scaleto{N,\omega}{5pt}}]^{\mathrm{T}}$, and (iii) the group with both excitations in the waveguide $c_{\omega_{1},\omega_{2}}$. $\bm{c}_{\scaleto{\mathrm{XX} }{4pt}}$ and $\bm{c}_{\scaleto{\mathrm{X\omega} }{4pt}}$ possess dimensions of $N_{\scaleto{\mathrm{XX} }{4pt}}=(N+1)(N+2)/2-1$ and $N+1$, respectively. As discussed in the previous section, we set the two-photon transport case with the intermediate state ($t = 0$) corresponding to group (i), which implies $\bm{c}_{\scaleto{\mathrm{X\omega} }{4pt}}(0) = 0$ and $c_{\omega_{1},\omega_{2}}(0) = 0$. The state reads
\begin{equation}
	\begin{aligned}\label{eqpsi0}
		\left|\Psi_{2}(0)\right\rangle=& \left[\sum_{n=0}^{N} \sum_{m > n}^{N} c_{n,m}(0) \hat{a}_{n}^{\dagger} \hat{a}_{m}^{\dagger} +\sum_{n=1}^{N} c_{n,n}(0) \frac{\hat{a}_{n}^{\dagger} \hat{a}_{n}^{\dagger}}{\sqrt{2}}\right]|\oslash\rangle.
	\end{aligned}
\end{equation}
Thus the equations of motion for the coefficients are \cite{Supp}
\begin{equation}\label{eq:CXX}
	\partial_{t} \bm{c}_{_{\scaleto{\mathrm{XX} }{4pt}}}=-i\mathbf{H}_{\scaleto{\mathrm{XX} }{4pt}}\bm{c}_{\scaleto{\mathrm{XX} }{4pt}},
\end{equation}
\begin{equation}\label{eq:CXw}
	\partial_{t} \bm{c}_{\scaleto{\mathrm{X\omega} }{4pt}}=-i(\mathbf{H}_{\scaleto{\mathrm{X} }{4pt}}+\mathbf{I}\scaleto{\Delta}{4.5pt}\omega)\bm{c}_{\scaleto{\mathrm{X\omega} }{4pt}}+\bm{d}_{\scaleto{\mathrm{X\omega} }{4pt}},
\end{equation}
\begin{equation}\label{eqCw1w2}
	\begin{aligned}
		\partial_{t} c_{\omega_{1},\omega_{2}}&=-i(\scaleto{\Delta}{4.5pt}\omega_{1}+\scaleto{\Delta}{4.5pt}\omega_{2})c_{\omega_{1},\omega_{2}}-i\sqrt{\frac{\kappa}{4\pi}}(c_{\scaleto{N }{4pt},\omega_{1}}+c_{\scaleto{N }{4pt},\omega_{2}}).
	\end{aligned}
\end{equation}
Here $\mathbf{H}_{\scaleto{\mathrm{XX} }{4pt}},\mathbf{H}_{\scaleto{\mathrm{X} }{4pt}}$ are the time-independent coupling matrices for $\bm{c}_{\scaleto{\mathrm{XX} }{4pt}}$ and $\bm{c}_{\scaleto{\mathrm{X\omega} }{4pt}}$ \cite{Supp}, respectively. $\scaleto{\Delta}{4.5pt}\omega_{1,2}=\omega_{1,2}-\omega_{0}$ is the relative detuning of photons with frequency $\omega_{1,2}$ from TLE.  The solution of Eq. \eqref{eq:CXX} enters Eq. \eqref{eq:CXw} through the driving term  $\bm{d}_{\scaleto{\mathrm{X\omega} }{4pt}}=\sqrt{\kappa/2/\pi}\allowbreak[c_{\scaleto{0,N}{5pt}},\allowbreak c_{\scaleto{1,N}{5pt}},....,c_{\scaleto{N-1,N}{5pt}},\allowbreak\sqrt{2}c_{\scaleto{N,N}{5pt}}]^{\text{T}}$. Sequentially, the time evolution of $c_{\omega_{1},\omega_{2}}$ is obtained from Eq. \eqref{eqCw1w2} based on the results of Eq. \eqref{eq:CXw}. By Fourier transforming $c_{\omega_{1},\omega_{2}}$ at a time sufficiently away from $t = 0$, we could analytically obtain the input and output two-photon wavepackets. In particular, the output wavepacket can be expressed as \cite{Supp}
\begin{equation}
	\begin{aligned}
		&g_{\mathrm{II}}\left(\tau_{1}, \tau_{2}\right)=-i \sqrt{\kappa \pi}  \sum_{l=1}^{N+1} \sum_{p=1}^{N_{\mathrm{xx}}}\xi_{l,p} \left(e^{-i\left(\lambda_{l}\left(\tau_{1}-\tau_{2}\right)+\Lambda_{p} \tau_{2}\right)} \right.\\
		&\left. \Theta\left[\tau_{1}-\tau_{2}\right]+e^{-i\left(\Lambda_{p} \tau_{1}+\lambda_{l}\left(\tau_{2}-\tau_{1}\right)\right)} \Theta\left[\tau_{2}-\tau_{1}\right]\right) \Theta(\tau_{1})\Theta(\tau_{2}),
	\end{aligned}
\end{equation}
where $\Theta$ is the Heaviside step function, $\Lambda_{n}$ and $\lambda_{n}$ are the eigenvalues of the matrices $\mathbf{H}_{\scaleto{\mathrm{XX} }{4pt}}$ and $\mathbf{H}_{\scaleto{\mathrm{X} }{4pt}}$, respectively. $\mathbf{I}$ is the identity matrix and $\xi_{l,p}$ are the coefficients determined from $|\Psi_{2}(0)\rangle$ and the eigenvectors of $\mathbf{H}_{\text{XX}}$ and $\mathbf{H}_{\text{X}} $. Such an analytical solution greatly facilitates us to get the optimal design of the gate. 

\par\emph{\textbf{Setting $|\Psi_{2}(0)\rangle$ for $\pi$-phase shift.---}} In this section, we discuss how to further narrow down the settings of $|\Psi_{2}(0)\rangle$ for acquiring a $\pi$ phase shift. We require that $|\Psi_{2}(0) \rangle$ is the system state at the time-reversal symmetric point, which means
\begin{equation}\label{eqpsit0relation}
	|\Psi_{2}(0)\rangle=|\Psi_{2}^{*}(0)\rangle, c_{n,m}(0) = c_{n,m}^{*}(0).
\end{equation}
Due to time-reversal symmetry, the phases acquired by the wavepacket during the absorption and emission processes are the same. The above scenario also applies to single photons. Let us examine the emission process of one excitation at $t = 0$. In our chain-coupled cavities and TLE, when an excitation hops from cavity $n$ (including the TLE labeled as $n=0$) to the neighboring cavity $n+1$ (or to the output waveguide), it acquires a phase of $\pi/2$. Thus, for the single-photon transport process, if the excitation at $t = 0$ resides in the n-th cavity $c_{n}(0) = 1$, then the single-photon wavepacket will acquire in total a phase of $(N-n+1)\pi$. For the two-photon transport process, if the two excitations at $t = 0$ reside in the n-th and m-th cavity, then two-photon wavepacket will acquire a phase of $(2N-n-m+2)\pi$. Thus, to obtain an extra $\pi$ phase shift compared to two single-photon transport processes, we have $c_{n,n+k}(0) = 0$ for even $k$.
\begin{table}[!b]
	\caption {Optimized gate parameters and fidelity ($g=1$)}\label{Table1}
	\begin{tabular}{|l|l|l|l|l|}
		\Xhline{1.5pt}
		$N$ 																			  & 1                                         & 2                                         & 3                                                                        & 4                                                                                      \\ \Xhline{1.3pt} 
		$\mathcal{P}_{1}$    													      & $\kappa=1.97$                                   &  \begin{tabular}[t]{@{}l@{}} $\kappa=3.62$\\ $J_{12}=1.52$  \end{tabular}                             &\begin{tabular}[t]{@{}l@{}}
			$\kappa=5.47$\\	$J_{12}=1.33$\\ $J_{23}=2.05$ \end{tabular}                                                           & \begin{tabular}[t]{@{}l@{}}
			$\kappa=7.43$\\
			$J_{12}=1.34$\\ $J_{23}=1.62$\\$J_{34}=2.60$\end{tabular} \\ \hline
		$\mathcal{P}_{2}$    										      & $c_{0,1}=1$                                         &\begin{tabular}[t]{@{}l@{}} $c_{0,1}=0.982$,\\$c_{1,2}=-0.191$\end{tabular}                            & \begin{tabular}[t]{@{}l@{}}$c_{0,1}=0.968$\\$c_{0,3}=-0.136$\\$c_{1,2}=-0.205$\\ $c_{2,3}=0.0433$\end{tabular} & \begin{tabular}[t]{@{}l@{}}$c_{0,1}=0.967$\\$c_{0,3}=-0.137$\\
			$c_{1,2}=-0.205$\\$c_{1,4}=0.0372$\\
			$c_{2,3}=0.0494$\\$c_{3,4}=-0.0118$\end{tabular} \\ \Xhline{1.3pt}
		$F_{\text{1}}$    & 0.942 & 0.985	 & 0.995                                                     & 0.998                                                                   \\  \Xhline{1.3pt}
		$F_{\text{2}}$ 									      & 0.895                            & 0.970                           &   0.988                                                           & 0.991                                                                      \\ \Xhline{1.3pt}
		$F_{\text{CZ}}$ 									      & 0.895                            & 0.970                           &  0.988                                                           & 0.991                                                                      \\ \Xhline{1.3pt}
		
	\end{tabular}
\end{table}

\begin{figure*}[!th]
	\centering
	\includegraphics[width=156mm]{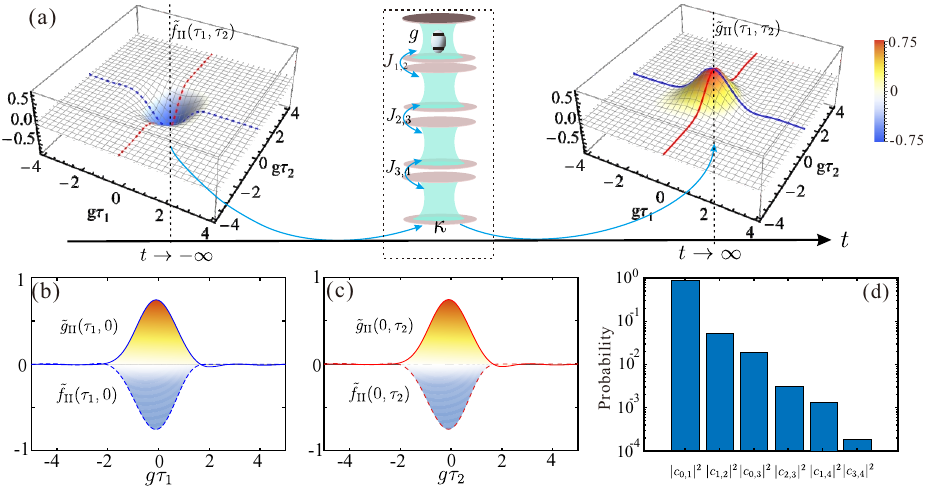}
	\caption{(a) Input and output two-photon wavepackets for the gate with $N=4$; Slices of the input (dashed line) and output (solid line) wavepackets at (b) $\tau_{2}=0$  (blue lines) and (c) $\tau_{1}=0$ (red lines). (d) Distribution of $|c_{n,n+k}(0)|^2$ ($k$ odd number).}
	\label{fig:fig2}
\end{figure*}
\par \emph{\textbf{High-fidelity for both single- and two-photon operations.---}} The gate fidelity is defined as the inner product of the ideal output state and the actual output after the gate operation \cite{NielsenBook2002quantum}. The wavepacket shape of the ideal output should be the same as the input. Therefore, the fidelities for the single-photon and two-photon cases can be respectively expressed as
\begin{equation} \label{eqF1}
	F_{1}=\left|\int \tilde{f}_{\mathrm{I}}^{*}(\tau)\tilde{g}_{\mathrm{I}}(\tau)d\tau\right|,
\end{equation}
\begin{equation} \label{eqF2}
	F_{2}=\left|\iint \tilde{f}_\mathrm{II}^{*}(\tau_{1},\tau_{2})\tilde{g}_{\mathrm{II}}(\tau_{1},\tau_{2})d\tau_{1}d\tau_{2}\right|.
\end{equation}
The functions with tilde are defined as $\tilde{f}_{\text{I}}(\tau)\equiv f_{\text{I}}(\tau-T/2),\tilde{g}_{\text{I}}(\tau)\equiv g_{\text{I}}(\tau+T/2)$ for single-photon wavepackets and $\tilde{f}_{\mathrm{II}}(\tau_{1},\tau_{2})\equiv f_{\mathrm{II}}(\tau_{1}-T/2, \tau_{2}-T/2),\quad\tilde{g}_{\mathrm{II}}(\tau_{1},\tau_{2})\equiv g_{\mathrm{II}}(\tau_{1}+T/2,\tau_{2}+T/2)$ for two-photon wavepackets, respectively. Here $f_{\text{I}/\text{II}}$ and $g_{\text{I}/\text{II}}$ are the input and output single-/two-photon wavepackets, respectively. The displacement $T/2$ is used to align the wavepackets in time for maximizing the overlap integrals in Eqs. \eqref{eqF1} and \eqref{eqF2}. As explained previously, $|\Psi_{1/2}(0)\rangle$ is the system state at the time-reversal symmetric point. Thus, for single-photon case, the input wavepacket has a time-reversed profile of the actual output, i.e., $i\tilde{f}_{\mathrm{I}}(\tau) =[i \tilde{g}_{\mathrm{I}}(-\tau)]^{*}$, which means that $F_{1}=|\int \tilde{g}_{\mathrm{I}}(-\tau)\tilde{g}_{\mathrm{I}}(\tau)d\tau|$.  The same argument also applies to the two-photon case with $F_{2} = |\iint \tilde{g}_{\mathrm{II}}(\tau_{1},\tau_{2})\tilde{g}_{\mathrm{II}}(-\tau_{1},-\tau_{2})d\tau_{1}d\tau_{2} |$. 
If the single-photon or two-photon wavepacket possesses a time-reversal-symmetric temporal shape, which gives  $i\tilde{g}_{\mathrm{I}}(-\tau)=[i\tilde{g}_{\mathrm{I}}(\tau)]^{*},\quad \tilde{g}_{\text{II}}(-\tau_{1},-\tau_{2})=[\tilde{g}_{\text{II}}(\tau_{1},\tau_{2})]^{*}$, then it can be shown $F_{1/2}=1$. 
Apart from that, there is an additional requirement for the two-photon state in a CZ gate, which demands that it should be the direct product state of the two single-photon states, so that $f_{\text{II}}(\tau_{1},\tau_{2})=f_{\text{I}}(\tau_{1})f_{\text{I}}(\tau_{2})$. This requirement leads us to define a cross symmetry factor for the wavepackets as
\begin{equation}\label{eqCrosssym}
	\mathcal{S}=\left|\iint \tilde{g}_{\mathrm{II}}(\tau_{1},\tau_{2}) \tilde{g}_{\mathrm{I}}(-\tau_{1})\tilde{g}_{\mathrm{I}}(-\tau_{2}) d\tau_{1}d\tau_{2} \right|.
\end{equation}

An analytical expression of $\mathcal{S}$ can be derived \cite{Supp}. It turns out that the optimization of $\mathcal{S}$ can simultaneously maximize the fidelity of both single- and two-photon operations.  The parameters to be optimized consists of two groups, i.e., (i) $\mathcal{P}_{1}=\left\{J_{1,2}/g,...,J_{N-1,N}/g,\kappa/g\right\}$ represents the coupling rates to characterize the system Hamiltonian, and (ii) $\mathcal{P}_{2}=\{c_{n,n+k}(0)|k \text{ is odd}\}$, with $\sum|c_{n,n+k}(0)|^2=1$ are the setting of the intermediate state $|\Psi_{2}(0)\rangle$. We have set $g = 1$ for convenience. For the optimization of these parameters, we apply the interior point method \cite{Supp}. As we show shortly, by maximizing  $\mathcal{S}$, we will obtain the optimal setting for the coupling rates $\mathcal{P}_{1}$, and the appropriate input single-photon state $|1\rangle=\int d\tau f_{\mathrm{I}}(\tau)\hat{c}_{\tau}^{\dagger}|\varnothing\rangle$ and the corresponding two-photon state $|2\rangle=\iint\allowbreak d\tau_{1}d\tau_{2} \allowbreak f_{\mathrm{I}}(\tau_{1}) \allowbreak f_{\mathrm{I}}(\tau_{2})(\hat{c}_{\tau_{1}}^{\dagger}\hat{c}_{\tau_{2}}^{\dagger}/\sqrt{2})|\varnothing\rangle$.

\par \emph{\textbf{Results and discussion.---}} We have studied the gate with cavities numbers of $N = 1\sim 4$ and presented the optimized parameters $\{\mathcal{P}_{1},\mathcal{P}_{2}\}$ and fidelities in Table I. The two-photon input state is formed by the product state of two identical single-photon states, which is obtained from the evolution of $|\Psi_{1}(0)\rangle$ according to Eqs. \eqref{eq:phi_t1} with the optimized $\mathcal{P}_{1}$. We found that the fidelity of both single- and two-photon cases gets improved as $N$ increases. In particular, for $N=4$, one could simultaneously obtain gate fidelities of over 99\% for both single-photon and two-photon cases. The CZ gate fidelity $F_{\text{CZ}}$ is defined as the minimum fidelity over all possible input states. In our scheme since we always have $F_{1}>F_{2}$, the CZ gate fidelity is determined by $F_{2}$.

We examine in detail the case of $N = 4$ with an optimized Hamiltonian. Figure \ref{fig:fig2}(a) displays the input and output two-photon wavepackets, respectively. The output wavepacket clearly shows a nonlinear $\pi$ phase shift as required by a CZ gate. Figure \ref{fig:fig2}(b) and \ref{fig:fig2}(c) are the slice cuts of the wavepackets along $\tau_{1}=0$ and $\tau_{2}=0$, respectively. One sees that, the temporal shapes of the input and output wavepackets are almost identical and are both time-reversal symmetric. Figure \ref{fig:fig2}(d) shows the probabilities $|c_{n,n+k}|^2$ at time $t=0$ with $k$ an odd number, which are the terms which could provide a nonlinear $\pi$ phase shift. The dominant term $|c_{0,1}|^2$ represents that at time $t = 0$ the two excitations reside in the TLE and its directly coupled cavity, consistent with the physical picture sketched in Fig. \ref{fig:fig1}. The full time evolutions of single-photon and two-photon transport processes are numerically calculated and support the above analysis \cite{Supp}. Moreover, the proposed gate also works well for single-photon and two-photon wavepackets with a Gaussian profile with an appropriate linewidth \cite{Supp}. The gate operation time is the order of $\sim 1/g$ and thus could be $\sim$1 ns with a GHz coupling rate. 

Next we discuss the prospect of experimental realization. The proposed gate architecture is compatible with integrated photonic platforms \cite{AOPReview2021,KippenbergOptica2018,LipsonLowLossResonator} and recent development in quantum photonics \cite{QianPhotoniX2021,SchmiedmayerPNAS2015,GreffetACSPho2017,ReisererRMPhy2022,WaksOptica2020,ZwillerNatPho2020,AharonovichNatPhon2016}. There are a number of excellent candidates of solid-state emitters such as quantum dot \cite{StobbeRevModPhys2015,WangNatPhon2019,AndrewNatNanotec2017}, color centers in diamond \cite{DirkNature2020} and molecules in organic crystals \cite{VahidNanoLett2017Molecu,RenNatComm2022,OrritNatMat2021} that have been demonstrated to be integrated with photonic circuits and show lifetime-limited linewdith at cryogenic temperatures. Moreover, a variety of platforms such as silicon nitride, lithium niobate and aluminium nitride \cite{LipsonLowLossResonator,AOPReview2021,HongOptica2018} have been shown for fabricating photonic microcavities with ultrahigh quality factors (small $\Gamma_{\text{C}}$). For quantum emitters with lifetime-limited linewidth of a few tens of MHz ($\Gamma_{0}$), emitter-cavity coupling constant $g$ of a few GHz, cavity-cavity and cavity-waveguide coupling rates of a few tens of GHz are achievable with realistic parameters. These properties could meet the preconditions of our proposal, which requires $g\gg\Gamma_{0}$, and $(J_{n,n+1}, \kappa) \gg \Gamma_{\text{C}}$. The possible mismatch between the resonances of the emitter and cavities could be minimized through various means, for instance Stark tuning of the emitter, post-fabrication trimming and MEMS actuation of the cavity resonances \cite{GleasonOL2005,JelenaAPL2008}. Our calculations show that the influences to the gate fidelity due to small $\Gamma_{0},\Gamma_{\text{C}}$ and detuning are manageable \cite{Supp}. 

In summary, we have demonstrated that a simple node, consisting of one quantum emitter and a small number of cavities, could realize a CZ gate for traveling single-photon qubits with a fidelity exceeding 99\%, surpassing the fault-tolerance threshold for scalable quantum computation \cite{DavidPRA2011ThresholdofFTQM}. The construction and operation of the gate are based on the principle of realizing time-reversal symmetric transport for both single- and two-photon wavepackets, which could simplify the gate operation to a process of perfect absorption and re-emission of the wavepackets by the node. Through this process, the quantum emitter induces a nonlinear $\pi$ phase shift for the two-photon wavepacket without distortion. The gate is deterministic and passive, significantly reducing the resource demand and the complexity of the operation. Further, the gate architecture is compatible with integrated photonic platforms, paving the way for scalable deterministic photonic quantum information processing.

\begin{acknowledgments} 
	\emph{\textbf{Acknowledgements.---}} We thank E. V. Stolyarov, Pu Zhang and Daiqin Su for fruitful discussions. This work is partially supported by the National Natural Science Foundation of China (Grant No. 92150111, 62235006), the Science and Technology Department of Hubei Province, China (Project No. 2022BAA018) and Huazhong University of Science and Technology.
\end{acknowledgments}
\bibliographystyle{apsrev4-1}
%

\end{document}